\documentstyle[aps]{revtex}
\begin{document}
%
\title{Invariant Box--Parameterization of Neutrino Oscillations}
\author{Thomas J. Weiler$^1$ and DJ Wagner$^{2}$}
\address{$^1$Department of Physics and Astronomy, Vanderbilt University,
   Nashville, TN  37235\\
$^2$Department of Physics and Astronomy, Angelo State University,
   San Angelo, TX  76909}
\maketitle
\begin{abstract}
The model-independent ``box" parameterization of neutrino oscillations is
examined.  The invariant boxes are the classical amplitudes of the
individual oscillating terms.  Being observables, the boxes are independent
of the choice of parameterization of the mixing matrix.
Emphasis is placed on the relations among
the box parameters due to mixing--matrix unitarity,
and on the reduction of the number of boxes to the minimum basis set.
Using the box algebra, we show
that CP-violation may be inferred from measurements
of neutrino flavor mixing even when the oscillatory factors have averaged.
General analyses of neutrino oscillations among $n\ge 3$ flavors
can readily determine the boxes,
which can then be manipulated to yield magnitudes of mixing matrix elements.
\end{abstract}
%
\def\beq{\begin{equation}}
\def\eeq{\end{equation}}
\def\bea{\begin{eqnarray}}
\def\eea{\end{eqnarray}}
\def\ba{\begin{array}}
\def\ea{\end{array}}
\def\ab{{\alpha \beta}}
\def\acb{{\alpha, \beta}}
\def\gsim{\raisebox{-0.5ex}{$\stackrel{>}{\sim}$}}
\def\lsim{\raisebox{-0.5ex}{$\stackrel{<}{\sim}$}}
\def\half{{\frac{1}{2}}}
\def\openone{\leavevmode\!\!\!\!\hbox{\small1\kern-1.55ex\normalsize1}}
\def\nue{{\nu_{e}}}
\def\numu{{\nu_{\mu}}}
\def\nutau{{\nu_{\tau}}}
\def\nui{{\nu_{i}}}
\def\nua{{\nu_\alpha}}
\def\nub{{\nu_\beta}}
\def\Pab{{P\makebox[9 mm][r]{\raisebox{-1.5ex}{{\scriptsize{$\nua \rightarrow
\nub$}}}}} \!\!\!\!\!\!\!\!\!\!\!(x)\;}
\def\Paa{{P\makebox[9 mm][r]{\raisebox{-1.5ex}{{\scriptsize{$\nua \rightarrow
\nua$}}}}} \!\!\!\!\!\!\!\!\!\!\!(x)\;}
\def\Pba{{P\makebox[9 mm][r]{\raisebox{-1.5ex}{{\scriptsize{$\nub \rightarrow
\nua$}}}}} \!\!\!\!\!\!\!\!\!(x)\;}
\def\Peu{{P\makebox[8 mm][r]{\raisebox{-1.5ex}{{\scriptsize{$\nue \rightarrow
\numu$}}}}} \!\!\!\!\!\!\!\!\!(x)\;}
\def\Put{{P\makebox[8 mm][r]{\raisebox{-1.5ex}{{\scriptsize{$\numu \rightarrow
\nutau$}}}}} \!\!\!\!\!\!\!\!\!(x)\;}
\def\Pet{{P\makebox[8 mm][r]{\raisebox{-1.5ex}{{\scriptsize{$\nue \rightarrow
\nutau$}}}}} \!\!\!\!\!\!\!\!\!(x)\;}
\def\Pabbar{{P\makebox[9 mm][r]{\raisebox{-1.5ex}{{\scriptsize{
$\overline{\nu}_{\alpha} \rightarrow \overline{\nu}_{\beta}$}}}}}
\!\!\!\!\!\!\!\!\!(x)\;}
\def\Pbabar{{P\makebox[9 mm][r]{\raisebox{-1.5ex}{{\scriptsize{
$\overline{\nu}_{\beta} \rightarrow \overline{\nu}_{\alpha}$}}}}}
\!\!\!\!\!\!\!\!\!(x)\;}
\newcommand{\Pnu}[2]{{P\!\!\!\!\makebox[9 mm][r]{\raisebox{-0.8ex}
  {{\tiny{${#1} \!\!\rightarrow \!\!{#2}$}}}}}}
\newcommand{\B}[2]{\;^{#1}\Box_{#2}\;}
\newcommand{\Bs}[2]{\;^{#1}\Box_{#2}^*\;}
\newcommand{\Baibj}{\B{\alpha i}{\beta j}}
\newcommand{\Bsaibj}{\Bs{\alpha i}{\beta j}}
\newcommand{\R}[2]{\;^{#1}\mbox{R}_{#2}}
\newcommand{\J}[2]{\;^{#1}\mbox{J}_{#2}}
\newcommand{\Raibj}{\R{\alpha i}{\beta j}}
\newcommand{\Jaibj}{\J{\alpha i}{\beta j}}
\def\Pij{{\Phi_{ij}}}
\def\Pji{{\Phi_{ji}}}
\def\Pe{{\Phi_{23}}}
\def\Pu{{\Phi_{13}}}
\def\Pt{{\Phi_{12}}}
\newcommand {\V}[1] {V_{#1}}
\newcommand {\Vs}[1] {V^*_{#1}}
\def\Vai{{V_{\alpha i}}}
\def\Vajs{{V_{\alpha j}^*}}
\def\Vbj{{V_{\beta j}}}
\def\Vbis{{V_{\beta i}^*}}
\newcommand {\co}[1] {c_{#1}}
\newcommand {\s}[1] {s_{#1}}
\newcommand {\ct}[1] {c^2_{#1}}
\newcommand {\st} [1] {s^2_{#1}}
\def\ssp{\hspace{0.3 in}}
\def\vsp{\vspace{0.5 cm}}
\newcommand {\combin}[2]
  {\left( \stackrel{\raisebox{0.8ex}{$\displaystyle {#1}$}}
  {\raisebox{-1.3ex}{$\displaystyle {#2}$}} \right)}
%

\section{Introduction}

If neutrinos have mass and are non-degenerate, then
their flavors may oscillate as they propagate.
Resonant oscillations for the sun\cite{solar},
oscillations for the atmosphere\cite{atmos},
and the LSND data\cite{LSND} each require a different
neutrino mass-squared difference if neutrino oscillations are to
account for all features of the data \cite{BWW}.
Since three-neutrino models can have at most two independent
mass-squared differences, a sterile neutrino is apparently needed to reconcile
all the data while retaining consistency with LEP measurements of
$Z \rightarrow \nu \overline{\nu}$ \cite{LEP}.  Several four-neutrino analyses
appear in the literature \cite{BWW,models}.  It is also possible that
some data will turn out to have an explanation other than neutrino oscillations,
in which case three-neutrino oscillations may be sufficient.
So our task is to examine the physics of neutrino oscillations with
three or more mixed flavors.

Oscillation probabilities depend on products of four mixing-matrix elements.
Several parameterizations of the mixing matrix in terms
of rotation angles have been introduced, beginning with the pioneering work of
Kobayashi and Maskawa \cite{KM}.  With three or
more neutrino generations, the oscillation
probabilities are complicated functions of the
neutrino mixing angles.  But oscillations are
observable and therefore parameterization-invariant.  One must ask if there is
not a better description of oscillations which avoids the arbitrariness of
angular-parameterization schemes.  Recently, we introduced a ``box''
parameterization of neutrino
mixing valid for any number of neutrino generations\cite{ww98}.
Oscillation probabilities are linear in the boxes, enabling a straighforward
description of oscillation data.  Here we
present the algebra of the boxes and the unitarity constraints on that algebra.
Then we illustrate the boxes' reduction to a basis in the case of three
generations, thereby setting the framework for a future
phenomenological analysis.




The probability for a neutrino to oscillate from $\nua$ to $\nub$ is given by
the square of the transition amplitude:
\beq
\Pab =  \left|\sum_{i=1}^{n} \Vai {\Vbis}
e^{-i \phi_i} \right|^2 =
\sum_{i=1}^{n} \sum_{j=1}^{n} (\Vai \;
\Vbis \; \Vajs \; \Vbj ) e^{-i \left( 2\Pij\right)},
\label{Probab}
\eeq
where $n$ is the number of neutrino generations,
\beq
\Phi_{ij} \equiv \half \left( \phi_i - \phi_j \right) = \half \left(
E_i t_i - p_i x_i - E_j t_j + p_j x_j \right),
\eeq
and $\Vai$ is the mixing--matrix element
which connects the $\alpha^{\rm th}$ charged lepton mass eigenstate and the
$i^{\rm th}$ neutrino mass eigenstate.
For relativistic neutrinos, $\Phi_{ij}$ is given by
\beq
\Phi_{ij} \approx \frac{\Delta m_{ij}^2}{4p} x, \mbox{\ \ where \ \ }
\Delta m_{ij}^2 \equiv m_i^2-m_j^2.
\label{Phiij}
\eeq
With a little bit of algebra, the oscillation probability
may be brought into the form
\bea
\Pab & = & -2 \sum_i \sum_{j \neq i} \mbox{Re}(\Vai \Vbis \Vajs \;\Vbj)
\sin^2 \left( \Phi_{ij} \right)
\label{oscillation}  \\
&&+ \ \sum_i \sum_{j \neq i} \mbox{Im}(\Vai \Vbis \Vajs \;\Vbj)
\sin \left( 2 \Phi_{ij} \right) + \delta_{\ab}.
\nonumber
\eea
The probability for an antineutrino to oscillate from $\overline{\nu}_{\alpha}$
to $\overline{\nu}_{\beta}$ is obtained by replacing
$V$ with $V^*$.  This is
equivalent to changing the sign of $\Pij$, or
the second term in equation~(\ref{oscillation}).

With the familiar case of two neutrino flavors, the
mixing matrix V  has the simple form of a rotation matrix
(phases cancel in oscillation probabilities for Majorana
neutrinos, and may be absorbed into the definitions of
Dirac fermion fields):
\beq
V = \left(
\begin{array}{cc}
       \mbox{cos} \,\theta    &   -\mbox{sin} \,\theta  \\
       \mbox{sin} \,\theta    &    \mbox{cos} \,\theta
\end{array}
\right).
\label{twoflavorV}
\eeq
The oscillation probability
in the two-flavor case is simply
\beq
\Pab = \delta_{\ab} + \mbox{sin}^2 \,2 \theta \;
\sin^2 \left( \frac{\Delta m^{2}_{12}}{4 p} x \right), \mbox{\ \ \ }n=2.
\label{twoflavorP}
\eeq
The mixing-angle parameterization is a natural
choice in the two-flavor situation.

The formalism becomes more complicated with three flavors.
An arbitrary $3 \times 3$ unitary matrix has three real
degrees of freedom and six phases, but
$2n-1=5$ phases may be absorbed into field redefinitions.
The original choice of the four remaining parameters,
due Kobayashi and Maskawa to describe quark mixing, is \cite{KM}
\beq
\left( \ba{ccc}
\co{1} & \;\s{1} \co{3}\; & \s{1} \s{3} \\
-\s{1}\co{2} & \; \co{1}\co{2}\co{3}-\s{2}\s{3}e^{i\delta}\; &
\co{1}\co{2}\s{3}+\s{2}\co{3}e^{i\delta} \\
-\s{1}\s{2} & \co{1}\s{2}\co{3}+\co{2}\s{3}e^{i\delta} &
\co{1}\s{2}\s{3}-\co{2}\co{3}e^{i\delta}  \ea \right),
\label{NactCKM}
\eeq
where $\co{a}\equiv \cos \theta_a$, and $\s{a} \equiv \sin \theta_a$.
There is arbitrariness associated with the
placement of the phase, since we absorb five relative phases into the field
definitions.  Because of this arbitrariness, the phases of individual
matrix elements are not observable.

The observable oscillation probabilities
are quite complicated functions of the angle-based
parameterizations.  As an example, consider the product
$\V{22}\Vs{23}\Vs{32}\V{33}$ appearing in the $\numu \rightarrow \nutau$
oscillation probability:
\bea
\V{22}\Vs{23}\Vs{32}\V{33} &=& \ct{3} \st{3} \left[ \st{2} \ct{2} (s_{1}^4 + 6
\ct{1} + 2 \ct{1} \cos 2 \delta) - \ct{1} \right]
\nonumber \\
 & &  + \frac{{\cal J}}{\st{1}} (1+\ct{1})(\ct{2}-\st{2})(\st{3} -
\ct{3}) \cot \delta + i {\cal J}, \mbox{\ \ \ }n=3.
\label{stanbox}
\eea
where the Jarlskog invariant ${\cal J}$ \cite{Jarls1} has the form
${\cal J} = \co{1} \st{1} \co{2} \s{2} \co{3} \s{3} \sin {\delta}$
in this parameterization.

The expression (but not its value)
on the right-hand side of equation (\ref{stanbox}) is convention-dependent,
as well as being unwieldy.
Our development of a model-independent parameterization
is motivated by the arbitrariness and complexity of this traditional
approach.


\section{The Box Parameterization}
\label{boxes}

The immeasurability of the individual complex mixing--matrix elements
in the quark sector has been
addressed by numerous authors \cite{Jarls1,NP,Wu,BD,DDW}.
Measurable quantities include only the
magnitudes of mixing matrix elements, the products of four mixing-matrix
elements appearing in the oscillation probabilities, and
particular higher-order functions of mixing-matrix elements \cite{Wu,KS}.
As evidenced in equations~(\ref{Probab}) and
(\ref{oscillation}), neutrino oscillation probabilities depend linearly on
the fourth-order objects,
\beq
\Baibj \equiv \Vai V^{\dagger}_{i \beta} \Vbj V^{\dagger}_{j \alpha}
= \Vai \V{\alpha j}^* \V{\beta i}^* \Vbj,
\label{boxdef}
\eeq
which we call ``boxes'' since each contains as factors the
corners of a submatrix, or ``box," of the mixing matrix.
For example, the upper left $2\times 2$ submatrix elements produce the box
\beq
\B{11}{22} = \V{11} \V{12}^* \V{21}^* \V{22}.
\eeq
The name ``box'' also seems appropriate in light of the Feynman
box--diagram which describes the oscillation process.
Examination of equation~(\ref{boxdef})
reveals a few symmetries in the indexing:
\beq
\Baibj = \B{\beta j}{\alpha i} = \Bs{\beta i}{\alpha j} =
\Bs{\alpha j}{\beta i}.
\label{symmetries}
\eeq
If the order of either set of indices is reversed ({\it id est},
$j \leftrightarrow i$
or $\beta \leftrightarrow \alpha$), the box turns into its complex conjugate;
if both sets of indices are reversed, the box returns to its original value
\cite{NP}.
And if $V$ is replaced by $V^{\dagger}$, then
$\Baibj\rightarrow \Bs{i\alpha}{j\beta}$.

Boxes with $\alpha = \beta$ or
$i = j$, are real, given from equation~(\ref{boxdef}) as
\beq
\B{\alpha i}{\alpha j} = |\V{\alpha i}|^{2} |\V{\alpha j}|^{2},
\mbox{\ \ \ \ }
\B{\alpha i}{\beta i} = |\V{\alpha i}|^{2} |\V{\beta i}|^{2},
\mbox{\ \ and \ \ }
\B{\alpha i}{\alpha i} = |\Vai|^4.
\label{samei}
\eeq
We call boxes with one and two repeated indices ``singly-degenerate'' and
``doubly-degenerate,'' respectively.  Boxes with $\alpha \neq \beta$ and $i\neq
j$ are called ``nondegenerate''.
As can be seen from equation (\ref{oscillation}), singly-degenerate boxes
with repeated flavor indices enter into the formulae for flavor-conserving
survival probabilities, but not for flavor-changing transition
probabilities.  Degenerate boxes with repeated mass indices (including
the doubly-degenerate boxes) do not appear in
any oscillation formula.  Degenerate boxes may be expressed in terms of
the nondegenerate boxes, as will be shown shortly. This
possibility and the symmetries expressed in
equation~(\ref{symmetries}) allow us to express combinations of boxes in terms
of only the
nondegenerate ``ordered'' boxes for which $\alpha < \beta$ and
$i < j$.




Using the symmetries expressed in equation~(\ref{symmetries}),
the oscillation probabilities (\ref{oscillation}) in terms of boxes
become
\beq
\Pab  = \delta_{\alpha \beta} -2 \sum_{i=1}^{n} \sum_{j>i}
\left[ 2 \R{\alpha i}{\beta j}
\sin^2 \Pij - \J{\alpha i}{\beta j} \sin 2 \Pij \right] ,
\label{boxorig}
\eeq
where we have defined the shorthand
$\R{\alpha i}{\beta j} \equiv \mbox{Re}\left(\Baibj\right)$
and $\J{\alpha i}{\beta j} \equiv\mbox{Im}\left(\Baibj\right)$.
From equation~(\ref{symmetries}) we
deduce that the $J$s are antisymmetric in both flavor indices and mass indices;
$R$s are symmetric in both.
Survival probabilities $\Paa = 1 - \sum_{\beta \neq \alpha} \Pab$
are more simply expressed in terms of degenerate boxes,
or $|V|$s, rather than nondegenerate boxes.  From equations~(\ref{boxorig})
and (\ref{samei}), they are
\beq
\Paa =  1-4\sum_{i=1}^n \sum_{j>i} \B{\alpha i}{\alpha j} \sin^2\Pij =
1-4\sum_{i=1}^n \sum_{j>i} |\V{\alpha i}|^2 |\V{\alpha j}|^2 \sin^2\Pij.
\eeq
Interchanging $\alpha \leftrightarrow \beta$ in equation (\ref{boxorig}) gives
the time-reversed reactions $\Pba$:
\beq
\Pba = \delta_{\alpha \beta} -2 \sum_{i=1}^{n} \sum_{j>i} \left[ 2
\R{\alpha i}{\beta j}
\sin^2 \Pij + \J{\alpha i}{\beta j} \sin 2 \Pij \right]\,.
\eeq

Ignoring possible CP-violating phases in the mixing matrix,
the number of real parameters determining $V$ is the number of
rotational planes available in n--dimensions, $N\equiv \half n(n-1)$ .
Determining these $N$ parameters determines the complete mixing matrix.
Conveniently, there are $N$ transition probabilities
$\Pab=\Pba$.
Thus, all of the information in the mixing matrix is contained in the $N$
transition probabilities.  In this sense, they form a convenient basis for
determining all oscillation parameters.  Of course, if the same transition
probability is measured at two or more different distances, then all $N$
transition probabilities may not be needed to determine $V$.

Allowing CP-violation in the mixing matrix, there are $N$ real parameters and
$\half (n-1)(n-2)$ phases, for a total of $(n-1)^2$ parameters.  With
CP-violation, however, there are $2N=n(n-1)$ independent transition
probabilities $\Pab$.  The number of transition probabilities exceeds the
number of independent parameters, so they again form a convenient
basis for determining the mixing matrix.
In reality, only the three flavor indices $e$, $\mu$, $\tau$ are easily
accessible.  Moreover, some
of the $N$ parameters in the mixing matrix, namely those which rotate sterile
states for $n\ge 5$, are not accessible at all, which complicates
the counting.

The transition probabilities for which $\alpha \neq \beta$ in
equation (\ref{boxorig}) may be conveniently expressed in matrix form.
The matrix of boxes is an $N\times N$ matrix.
For three flavors, we have
\beq
{\cal P}(n=3) \equiv
\left( \ba{c}
\Peu \\ \Put \\ \Pet
\ea \right) =
-4 \mbox{ Re}({\cal B})\; S^2(\Phi) + 2 \mbox{ Im} ({\cal B})\; S(2\Phi),
\label{boxprob}
\eeq
where
\beq
{\cal B} \equiv
\left( \ba{ccc}
\B{e1}{\mu 2} & \B{e2}{\mu 3} & \B{e1}{\mu 3} \\
\B{\mu 1}{\tau 2} & \B{\mu 2}{\tau 3} & \B{\mu 1}{\tau 3} \\
\B{e1}{\tau 2} & \B{e2}{\tau 3} & \B{e1}{\tau 3}
\ea \right), \mbox{\ \ and \ \ }
S^k(\Phi) \equiv
\left( \ba{c}
\sin ^k\Pt \\  \sin^k \Pe \\ \sin^k \Pu
\ea \right), \mbox{\ \ \ }n=3.
\label{boxbox}
\eeq
For the time--reversed channels, or for the antineutrino channels, the sign of
the Im$({\cal B})$ term is reversed.  The box parameterization is
especially well-suited for considering higher numbers of generations.
The matrix ${\cal B}$ merely acquires extra
columns when new flavors are introduced; extra rows are not accessible at
energies below new charged-lepton thresholds.  Furthermore,
oscillation probabilities are linear in boxes, no matter how many generations.




Neutrino oscillation experiments will directly measure the boxes in equation
(\ref{boxorig}),
not the individual mixing matrix elements, $\Vai$.  But one would like to obtain
the fundamental $\Vai$ from the measured boxes.  We develop here an algebra
relating boxes and mixing matrix elements.

Some tautologous relationships
between the degenerate and nondegenerate boxes are easily confirmed using
equation  (\ref{boxdef}); they hold for any number of generations:
\bea
|V_{\alpha i}|^2 |V_{\alpha j}|^2 & = & \B{\alpha i}{\alpha j} =
   \frac{\Bs{\alpha i}{\eta j} \B{\alpha i}{\lambda j}}{\B{\eta i}{\lambda j}},
 \;\;\;\;\;\;\;\;\;\;\;\;
 (\eta \neq \lambda \neq \alpha),
 \label{vproda} \\
\nonumber \\
|V_{\alpha i}|^2 |V_{\beta i}|^2 & = & \B{\alpha i}{\beta i} =
  \frac{\Bs{\alpha i}{\beta x} \B{\alpha i}{\beta y}}{\B{\alpha x}{\beta y}},
 \;\;\;\;\;\;\;\;\;\;\;
 (x \neq y \neq i),  \mbox{\ \ and \ \ }
 \label{vprodi} \\
\nonumber \\
\frac{|\V{\alpha i}|^2}{|\Vbj|^2} & = &
   \frac{\Bs{\alpha i}{\eta j}\B{\alpha i}{\beta x}}
   {\B{\alpha j}{\beta x} \Bs{\beta i}{\eta j}},
 \;\;\;\;\;\;\;\;\;\;(\eta \neq \alpha \neq \beta,
 \mbox{ and } x \neq i \neq j).
 \label{vratio}
\eea

Equations (\ref{vproda}) and (\ref{vprodi})
are themselves special cases of the more general
\bea
\label{2boxij}
\Baibj \B{\gamma i}{\delta j} &=& \left[ \Vai \Vajs \Vbj \Vbis \right]
\left[ V_{\gamma i} \Vs{\gamma j} V_{\delta j} \Vs{\delta i} \right] \\
& = &\left[ \Vai \Vajs V_{\delta j} \Vs{\delta i}\right]
\left[ V_{\gamma i} \Vs{\gamma j} \Vbj \Vbis \right] =
\B{\alpha i}{\delta j} \B{\gamma i}{\beta j}, \nonumber
\eea
and the analogous relation
$\Baibj \B{\alpha k}{\beta l} = \B{\alpha i}{\beta l} \B{\alpha k}{\beta j}$.
The relations above hold for both degenerate boxes and nondegenerate boxes.

Due to the symmetry $\Baibj \rightarrow \Bs{i\alpha}{j\beta}$ when
$V\rightarrow V^{\dagger}$, there will generally be analogous but distinct
pairing of our box equations,
differing only in whether the degeneracy or sum is over a flavor index
or a mass index.
In the following we will mainly show only one equation per analogous pair,
for reasons of space limitations in this proceeding.

We may express $|V_{\alpha i}| = \left(\B{\alpha i}{\alpha
i}\right)^{\frac{1}{4}}$
in terms of three singly-degenerate boxes by setting $\alpha=\beta$
in equation~(\ref{vprodi}).  Then, using equation (\ref{vproda})
to substitute for the singly-degenerate boxes yields an expression for the
doubly-degenerate box in terms of nine nondegenerate boxes:
\beq
|V_{\alpha i}|^4 = \B{\alpha i}{\alpha i} =
\frac{\B{\alpha i}{\alpha x} \B{\alpha i}{\alpha y}}{\B{\alpha x}{\alpha y}} =
\frac{\B{\alpha x}{\tau i} \B{\alpha i}{\sigma x}  \B{\alpha y}{\rho i}
\B{\alpha i}{\zeta y} \B{\omega x}{\mu y}} {\B{\tau i}{\sigma x} \B{\rho
i}{\zeta y} \B{\alpha y}{\omega x}  \B{\alpha x}{\mu y}}\,,
\label{vfour}
\eeq
where the index constraints are
$\tau \neq \sigma \neq \alpha$, $\zeta \neq \rho \neq \alpha$,
$\mu \neq \omega \neq \alpha$, and $x \neq y \neq i$.
In the three-generation case, equation (\ref{vfour})
is uniquely specified by the index constraints.
For example,
\beq
|\V{11}|^4 =
\frac{\B{11}{22} \Bs{11}{23} \B{11}{33} \Bs{11}{32} \B{22}{33}}
{\Bs{12}{23} \B{12}{33} \B{21}{33} \Bs{21}{32}}
\label{V11four}
\eeq
holds with any number of generations, but it is the unique $5$
on $4$ box representation of $\left|\V{11}\right|^4$ in three generations.

We note that all of the relationships in this section
follow from the definitions of the
boxes in equation (\ref{boxdef}) and so are valid for any matrix, unitary or
otherwise.  The constraints of unitarity
will provide us with expressions for $|\Vai|^4$
which are easier to manage than the expression in (\ref{vfour}) above.


\section{Unitarity Relations Among the Boxes}


Unitarity requires  that
\beq
\sum_{\eta=1}^n \V{\eta i} \V{\eta j}^* = \delta_{ij},
 \mbox{ \ \ and \ \ }
\sum_{y=1}^n \V{\alpha y} \V{\beta y}^* = \delta_{\alpha \beta}.
\label{Vunit}
\eeq
Multiplying the first equation in (\ref{Vunit}) by
$V_{\lambda i}^* V_{\lambda j}$ and the second by
$V_{\alpha x}^* V_{\beta x}$ gives the unitarity constraints
for the boxes:
\beq
\hspace{1.0 cm} \sum_{\eta=1}^n \B{\eta i}{\lambda j}
  =\sqrt{\B{\lambda i}{\lambda i}} \delta_{ij},
\label{urow}
\eeq
and its analogue.
Isolating the manifestly degenerate boxes from the nondegenerate boxes,
equation~(\ref{urow}) becomes
\beq
\sum_{\eta \neq \lambda} \B{\eta i}{\lambda j} =
\sqrt{\B{\lambda i}{\lambda i}} \delta_{ij} - \B{\lambda i}{\lambda j}.
\label{prof1}
\eeq

Summing equation~(\ref{urow}) over $\lambda$ in the $i\neq j$ case, we find
\beq
0=\sum_{\lambda=1}^n \sum_{\eta=1}^n \B{\eta i}{\lambda j} =
\sum_{\lambda=1}^n \B{\lambda i}{\lambda j}
+ 2 \sum_{\lambda=1}^n \sum_{\eta < \lambda} \R{\eta i}{\lambda j}.
\label{sumrowB}
\eeq
The double sum is over $R$s only, since the first term is manifestly real.  The
resulting conditions on the $J$s are found in equation~(\ref{imaguni}) below.
Comparison of equation~(\ref{sumrowB}) with equations~(\ref{boxbox}) and
(\ref{vproda}) reveals an interesting property of the matrix ${\cal B}$:
\beq
\sum_{\mbox{\small{column}}\,of\,{\cal B}} \mbox{Re}\left({\cal B}\right) =
-\half \sum_{\lambda=1}^n |V_{\lambda i}|^2 |V_{\lambda j}|^2,
\label{hereiam}
\eeq
where the sum is over a column of ${\cal B}$ specified by fixed $i$ and $j$.
There is an analogue relation for the sum over a row of ${\cal B}$.

The unitarity constraint (\ref{urow})
holds independently for the real and imaginary parts of the sum.
We will first explore the implications of these constraints for the imaginary
parts of boxes, before
turning to the more complicated constraints for the real parts.
The right-hand side of (\ref{urow})  is manifestly real, so
the imaginary constraints are simply
\beq
\sum_{\eta \neq \lambda} \J{\eta i}{\lambda j} = 0,  \mbox{ \ \ and \ \ }
\sum_{y \neq x} \J{\alpha y}{\beta x}= 0\,.
\label{imaguni}
\eeq
Equation~(\ref{symmetries}) indicates that
$\J{\eta i}{\lambda j}$ is an antisymmetric matrix in the indices $\eta$ and
$\lambda$ for fixed $i$ and $j$, and vice versa.
Equation~(\ref{imaguni}) shows that the sum
of elements along any row or column of that antisymmetric matrix equals zero,
whether the sum is over mass indices or flavor indices.

Summing the first equation in (\ref{imaguni}) over $\lambda$ gives
zero trivially since
a sum of all elements of an antisymmetric matrix vanishes by definition.
Hence, for fixed $\left(i,j\right)$, the first equation in (\ref{imaguni})
expresses $n-1$ constraints.
Thus, the number of independent flavor pairs on $J$s
after implementing the constraints of equation~(\ref{imaguni}) is
$N-(n-1)=\half (n-1)(n-2)$. Ditto for independent mass pairs, so
the number of independent $J$s after implementing both sets of
constraints is the product
$\frac{1}{4}\left(n-1\right)^2\left(n-2\right)^2$.

In three generations, this number of independent $J$s is one.
Each sum in equation~(\ref{imaguni}) contains only two terms,
leading to
\beq
\mbox{Im} ({\cal B}) = \left( \ba{ccc}
{\cal J} &  {\cal J} & -{\cal J} \\
{\cal J} &  {\cal J} & -{\cal J} \\
-{\cal J} &  -{\cal J} & {\cal J}
\ea \right),
\mbox{\ \ \ $n=3$},
\label{allJ}
\eeq
with ${\cal J} \equiv \J{11}{22}$ \cite{Jarls1}.
One consequence of the equality of all $\left|J\right|$s 
in three generations is that if any one $\Vai$ is zero, then all 
$\J{\alpha i}{\beta j}$ vanish and there can be no CP-violation.

We now consider the real parts of the constraint (\ref{urow}),
focusing first on the homogeneous constraint
for which the Kronecker delta
is zero.
This constraint gives the singly-degenerate boxes as sums of
ordered boxes:
\beq
|V_{\lambda i}|^2|V_{\lambda j}|^2 =
\B{\lambda i}{\lambda j} = - \sum_{\eta \neq \lambda} \B{\eta i}{\lambda j}
=- \sum_{\eta \neq \lambda} \R{\eta i}{\lambda j},
\hspace{0.4 in} i \neq j.
\label{1degrow}\\
\eeq
This linear relation complements the relation expressed in
equation~(\ref{vproda}).
For three generations, each of the sums contains two terms, allowing us to
express the singly-degenerate boxes in terms of two nondegenerate boxes
which are measurable in neutrino appearance oscillation experiments.

The real unitarity constraint (\ref{1degrow})
greatly simplifies our expressions for a
doubly-degenerate box $\B{\alpha i}{\alpha i} = |\V{\alpha i}|^4$:
\beq
\B{\alpha i}{\alpha i} =
\frac{\B{\alpha i}{\alpha x} \B{\alpha i}{\alpha y}} {\B{\alpha x}{\alpha y}} =
\frac{\left(- \sum_{\eta \neq \alpha} \R{\alpha i}{\eta x} \right)
\left(- \sum_{\lambda \neq \alpha} \R{\alpha i}{\lambda y} \right)}
{\left(- \sum_{\tau \neq \alpha} \R{\alpha x}{\tau y} \right)},
\ssp x \neq y \neq i,
\label{2degxy}
\eeq
where the first equality is due to equation~(\ref{vproda})
with $j=i$.
Applying equation~(\ref{2degxy}) to three generations, one
finds that doubly-degenerate boxes are expressible
in terms of the real parts of
six ordered boxes, rather than the nine complex boxes
used in equation (\ref{vfour}).  For example,
\beq
|V_{11}|^4 = \B{11}{11} =
\frac{-\left(\R{11}{22}+\R{11}{32}\right) \left(\R{11}{23}+\R{11}{33}\right)}
{\R{12}{23}+\R{12}{33}}, \mbox{\ \ \ $n=3$}.
\label{B1111}
\eeq
In cyclic coordinates $\alpha, \beta, \gamma$, and with $x\neq y\neq i$,
\beq
\B{\alpha i}{\alpha i} = |\Vai|^4 =
\frac{-\left(\R{\alpha i}{\beta x}+\R{\alpha i}{\gamma x}\right)
\left(\R{\alpha i}{\beta y}+\R{\alpha i}{\gamma y}\right)}
{\R{\alpha x}{\beta y}+\R{\alpha x}{\gamma y}}, \mbox{\ \ \ }n=3.
\label{Bcyclic}
\eeq
When considering $n>3$, each sum has more terms, but all terms in the
numerator in equation
(\ref{2degxy}) always contain $R$s to the second order,
while the denominator terms contain only the first order of $R$s.  Thus these
expressions will be much more manageable than equation (\ref{vfour})
which exhibits the fifth order of complex boxes in the numerator and the
fourth order in the denominator.

For fixed $(i,j)$ in equation~(\ref{1degrow}), $\lambda$ can take $n$
possible values, implying $n$ constraint equations.  $N$
ordered nondegenerate boxes appear in these $n$ equations.
Thus, for $N\le n$, which is true for $n\le 3$,
the unitarity constraint (\ref{1degrow}) may be inverted to
find a nondegenerate box in terms of singly-degenerate boxes.
Manipulation of equation~(\ref{1degrow}) gives an
expression in term of the flavor triad $(\alpha, \beta, \gamma)$:
\beq
\R{\alpha i}{\beta j} = -\half \left(
|V_{\alpha i}|^2|V_{\alpha j}|^2 + |V_{\beta i}|^2|V_{\beta j}|^2 -
|V_{\gamma i}|^2|V_{\gamma j}|^2 \right), \mbox{\ \ \ }n=3.
\label{RKim2}
\eeq

It is known that knowledge of four $|V|$s completely specifies the
three-generation mixing matrix, provided no more than two $|V|$s are
taken from the same row or same column \cite{HnJ}.  Here, we can
use three-generation unitarity and
equation~(\ref{RKim2}) to re-write $\R{\alpha i}{\beta j}$
in terms of just four $|V|$s.  The result is
\beq
\R{\alpha i}{\beta j} = \half \left[1-\left|\Vai\right|^2 -
\left|\V{\alpha j}\right|^2 - \left|\V{\beta i}\right|^2 - \left|\Vbj\right|^2 +
\left|\Vai\right|^2\left|\Vbj\right|^2 +
\left|\V{\alpha j}\right|^2 \left|\V{\beta i}\right|^2 \right],
\label{RVs}
\eeq
which for $n=3$ expresses the real part of the box
Re$\left[\Vai\Vajs\Vbj\Vbis\right]$ in terms of the magnitudes of the four
complex $V$s which define the box.  Three-generation unitarity may be used
again to replace the first five terms on the right-hand side of
equation~(\ref{RVs}) with $-\left|V_{\gamma k}\right|^2$.

Summing equation~(\ref{1degrow}) over $j\neq i$ yields another
expression for $|V_{\lambda i}|^2$ in terms of nondegenerate boxes,
which further complements equations (\ref{2degxy}) and (\ref{vfour}):
\beq
|V_{\lambda i}|^2 \sum_{j\neq i} |V_{\lambda j}|^2 =
|V_{\lambda i}|^2 \left(1-|V_{\lambda i}|^2\right) =
-\sum_{j\neq i} \sum_{\eta \neq \lambda} \R{\eta i}{\lambda j}.
\eeq
The explicit solution of this equation, valid for any number of
generations, is,
\beq
|V_{\lambda i}|^2 = \half\left[
1 \pm \sqrt{1+4\sum_{j\neq i} \sum_{\eta \neq \lambda} \R{\eta i}{\lambda
j}}\right],
\label{profeq}
\eeq
which yields $|V_{\lambda i}|^2$ in terms of $(n-1)^2$ $R$s,
but subject to a two-fold ambiguity.

We may use the real homogeneous unitarity condition
(\ref{1degrow}) along with the tautology (\ref{vproda})
to obtain constraints between
nondegenerate boxes, thereby reducing the number of real degrees of freedom.
Substituting the tautology (\ref{vproda})
into the unitarity constraint (\ref{1degrow}) gives
\beq
\B{\eta i}{\alpha j} \B{\alpha i}{\lambda j} + \B{\eta i}{\lambda j}
  \sum_{\tau \neq \alpha} \R{\tau i}{\alpha j} = 0.
\label{prof75}
\eeq
This unitarity constraint interrelates imaginary and real parts of $n$ different
boxes for any number of generations.
For example, taking the imaginary part of equation~(\ref{prof75})
leads to
\beq
\J{\eta i}{\alpha j} \R{\alpha i}{\lambda j} +
\J{\alpha i}{\lambda j} \R{\eta i}{\alpha j} +
\J{\eta i}{\lambda j} \sum_{\tau \neq \alpha} \R{\tau i}{\alpha j} = 0,
\mbox{\ \ \ $\eta \neq\lambda\neq\alpha$, \ \ $i\neq j$}.
\label{prof77}
\eeq
Taking the real part of equation~(\ref{prof75}) leads to
\beq
\R{\eta i}{\alpha j} \R{\alpha i}{\lambda j} +
\R{\eta i}{\lambda j} \sum_{\tau \neq \alpha} \R{\tau i}{\alpha j} =
\J{\eta i}{\alpha j} \J{\alpha i}{\lambda j},
\mbox{\ \ \ $\eta \neq\lambda\neq\alpha$, \ \ $i\neq j$}.
\label{prof78}
\eeq
One may also use the pairs of equations~(\ref{prof77}) and
(\ref{prof78}) to eliminate the sums and isolate a single $R$:
\beq
\R{\alpha i}{\beta j} =
  \frac{\J{\alpha i}{\beta j} \R{\alpha i}{\lambda j} \R{\beta i}{\lambda j}
    + \J{\alpha i}{\beta j} \J{\alpha i}{\lambda j} \J{\beta i}{\lambda j}}
  {\J{\alpha i}{\lambda j} \R{\beta i}{\lambda j}
    - \R{\alpha i}{\lambda j} \J{\beta i}{\lambda j}}\,,
\label{realconstraint}
\eeq
with $\beta \neq \lambda \neq \alpha$, $i \neq j$.
Input from these unitarity relations among $R$s and $J$s is necessary
to establish the minimum set of independent box parameters.

\section{Indirect Measurement of CP-Violation}

Suppose CP is conserved.  Then $\J{\alpha i}{\beta j}=0$ for all index choices.
The inference from equation~(\ref{prof78}) is that
\beq
\R{\alpha i}{\eta j} \R{\alpha i}{\lambda j} + \R{\eta i}{\lambda j}
\sum_{\tau \neq \alpha} \R{\alpha i}{\tau j} = 0,
\hspace{0.5 in} (\eta \neq \lambda \neq \alpha, \mbox{\ and \ } i\neq j).
\eeq
If this relation is violated, then so is CP.

For three generations,
$\J{\eta i}{\alpha j}= \J{\alpha i}{\lambda j}$
by equation~(\ref{imaguni}), and
equation (\ref{prof78}) may be solved for ${\cal J}^2$ directly:
\beq
{\cal J}^2  =
\R{\alpha i}{\beta j} \R{\beta i}{\lambda j} +
\R{\alpha i}{\beta j} \R{\alpha i}{\lambda j} +
\R{\alpha i}{\lambda j} \R{\beta i}{\lambda j}, \mbox{\ \ \ }n=3.
\label{CPone}
\eeq
Equation~(\ref{CPone}) says that the three real
elements in any row (or any column in the analogue equation)
of the matrix ${\cal B}$ may be summed in their
three pairwise products to yield the CP-violating invariant ${\cal J}^2$.
These real elements on the right-hand side of this equation
are measurable with CP-conserving averaged neutrino oscillations.
Thus, even if CP violating asymmetries are not
directly observable in an experiment, the effects of CP violation may be seen
through the relationships among the real parts of different boxes,
which are determinable from averaged flavor--mixing measurements!
Note that if CP is conserved and ${\cal J}$ is zero, then
equation~(\ref{CPone}) also tells us that all three $R$s in any row (or column)
cannot have the same sign.

\section{Inhomogeneous Unitarity Constraints and a Box--Basis}
\label{imhogsec}

The inhomogeneous unitarity constraints with the Kronecker delta nonzero
in equation~(\ref{urow}) are necessary to provide
the desired normalization of the $V_{\alpha i}$ or the boxes.
The inhomogeneous constraints
are functions of degenerate boxes and therefore purely real:
\beq
\B{\alpha i}{\alpha i} + \sum_{\eta \neq \alpha} \B{\alpha i}{\eta i}=
\sqrt{\B{\alpha i}{\alpha i}},
\label{2degsuma}
\eeq
This equation
can be rewritten strictly in terms of nondegenerate boxes
by using the homogeneous unitarity constraints (\ref{1degrow})
to replace the singly-degenerate boxes, and
equation~(\ref{2degxy}) to replace the doubly-degenerate box:
%
%
\beq
\Sigma_{\lambda}\Sigma_{\sigma}
  +\sqrt{-\Sigma_{\lambda}\Sigma_{\sigma}\Sigma_{\tau}}
  +\Sigma_{\tau}\Sigma_{\eta z} = 0\,,
\label{tuffy}
\eeq
with
$\Sigma_{\lambda}\equiv \sum_{\lambda \neq \alpha} \R{\alpha i}{\lambda x}$,
$\Sigma_{\sigma}\equiv \sum_{\sigma \neq \alpha} \R{\alpha i}{\sigma y}$,
$\Sigma_{\tau}  \equiv \sum_{\tau \neq \alpha} \R{\alpha x}{\tau y}$,
$\Sigma_{\eta z}\equiv\sum_{\eta \neq \alpha}\sum_{z \neq i}\R{\alpha
z}{\eta i}$,
and $x \neq y \neq i$.
These inhomogeneous unitarity
constraints do not involve the $J$s.
Isolating the square root and squaring the equation,
we get polynomial equations of
degrees three and four in the $R$s, each relating
$n(n-1)$ $R$s.

We provide here an example of a basis construction,
obtained by substituting in the unitarity equations derived above.
The unitarity constraints among the $J$s, given in
equation~(\ref{imaguni}),
are linear and therefore the simplest to implement.
These constraints may be used first to reduce the number of
independent $J$s to $\frac{1}{4}\left(n-1\right)^2\left(n-2\right)^2$.
 Further reduction to
independent $J$s and $R$s requires the nonlinear constraints.
The homogeneous constraints~(\ref{prof77}) and
(\ref{prof78}) are much simpler than the
inhomogeneous constraints~(\ref{tuffy}),
but the inhomogeneous constraints must be invoked at least once
(Otherwise, the the boxes and the matrix
element magnitudes $|V|$ could not be normalized.).

For three generations, one begins with nine $R$s and one $J$,
and seeks a basis of just four elements.
Rearranging the three-generation equation~(\ref{CPone})
yields expressions for one $R$ in terms of two other $R$s and ${\cal J}$.
This equation may be used three times to
eliminate $\R{12}{23}$, $\R{11}{32}$ and $\R{21}{33}$.
The utility of the analogue equation is exhausted to eliminate
$\R{12}{33}$ and $\R{11}{33}$.
As expected, one must next turn to the inhomogeneous
constraints (\ref{tuffy}) to eliminate the last degree of freedom.
We are left with a constraint which is quartic in all five
of its parameters
$A\equiv\R{11}{22}$, $B\equiv\R{11}{23}$, $C\equiv\R{21}{32}$,
$D\equiv\R{22}{33}$, and ${\cal J}^2$:
\bea
0&=&\left(A+B\right)^2\left(A+C\right)^2\left(BC+BD-CD+{\cal J}^2\right)^2
\nonumber \\
&&+\left(AD+BD-AB+{\cal J}^2\right)^2
\left[\left(A+B\right)\left(C+D\right)+C^2+{\cal J}^2\right]^2
\label{quartic} \\
&& + \left(A+B\right)\left(A+C\right)\left(BC+BD-CD+{\cal J}^2\right)
\left(AD+BD-AB+{\cal J}^2\right)
\nonumber \\
&&\hspace{0.2 cm}\times \left[C+D+2\left(\left(A+B\right)\left(C+D\right)
+C^2+{\cal J}^2\right)\right].
\nonumber
\eea
We may eliminate any one parameter
by either algebraic or numerical means, leaving us with the desired
four parameters as the basis.

\section{summary}

Neutrino physics has entered a golden age of research.  New experiments all over
the globe promise an unequaled amount of data from the sun, the atmosphere,
accelerators, supernovae, and other cosmic sources.  The latest data suggests
that more than three neutrino flavors may participate in neutrino oscillations
\cite{BWW}.  Analyzing such refined data
requires a consistent, model-independent approach which may be easily applied,
and easily extended to higher generations.
Here we have discussed such an approach,
wherein one works directly with the observable coefficients of the
oscillating terms.
From unitarity of the mixing matrix, we derived relations among these
CP--conserving and CP--violating coefficients for the various oscillation
channels.
One result which we view as particularly noteworthy is that high-statistics
data on
{\em averaged} oscillations are sufficient to determine the conservation or
non-conservation of CP in the lepton mixing matrix.  This indirect test of CP
can be traced back to unitarity of the mixing matrix,
but in the present formulation there is no need to even mention the
mixing matrix.\\
{\bf Acknowledgements}:  This work was supported in part by the U.S.
Department of Energy, Division of High Energy Physics,
under Grant No. DE-F605-85ER40226,
and the Vanderbilt University Research Council.


\end{document}